\documentclass[journal]{IEEEtran}
\usepackage{times}
\usepackage{epsfig}
\usepackage{graphicx}
\usepackage{amsmath}
\usepackage{amssymb}
\usepackage{amsmath,graphicx}
\usepackage{amsmath,amssymb,amsfonts}
\usepackage[labelfont=bf,list=true,font=footnotesize, margin=2pt]{subcaption}
\usepackage[labelfont=bf, margin=0pt]{caption}
\usepackage{balance,cite}
\usepackage{algorithm}
\usepackage{algorithmic}
\usepackage{siunitx}
\usepackage{setspace,color, enumitem}
\usepackage{flushend}
\usepackage{sidecap}
\usepackage{float}
\usepackage{xr} 

\newcommand{\balpha}{\ensuremath{\boldsymbol{\alpha}}}

\usepackage{cite}

\ifCLASSINFOpdf
\else
\fi

\usepackage[normalem]{ulem}
\newif\ifannotated

\annotatedtrue
\annotatedfalse

\ifannotated

\newcommand{\delete}[1]{{\color{red}{\sout{#1}}}}

\newcommand{\margincomment}[1]{\marginpar{$\Rightarrow$\color{red}\fbox{\parbox{\linewidth}{\color{black}\scriptsize#1}}}}
\else

\newcommand{\delete}[1]{\ignorespaces}

\newcommand{\margincomment}[1]{}
\fi

\begin{document}

\title{Joint Image and Depth Estimation \\ with Mask-Based Lensless Cameras}

\author{Yucheng Zheng and M. Salman Asif
\thanks{
Y. Zheng and M. Asif are with the Department
of Electrical and Computer Engineering, University of California, Riverside,
CA, 92521 USA (e-mail: yzhen069@ucr.edu; sasif@ece.ucr.edu).}
\thanks{A shorter version of this paper with preliminary results was presented in \cite{zheng2019camsap}}

}

\maketitle

\begin{abstract}
    Mask-based lensless cameras replace the lens of a conventional camera with a custom mask. These cameras can potentially be very thin and even flexible. Recently, it has been demonstrated that such mask-based cameras can recover light intensity and depth information of a scene. Existing depth recovery algorithms either assume that the scene consists of a small number of depth planes or solve a sparse recovery problem over a large 3D volume. Both these approaches fail to recover the scenes with large depth variations. In this paper, we propose a new approach for depth estimation based on an alternating gradient descent algorithm that jointly estimates a continuous depth map and light distribution of the unknown scene from its lensless measurements. We present simulation results on image and depth reconstruction for a variety of 3D test scenes. A comparison between the proposed algorithm and other method shows that our algorithm is more robust for natural scenes with a large range of depths. We built a prototype lensless camera and present experimental results for reconstruction of intensity and depth maps of different real objects.  
\end{abstract}


%
\IEEEpeerreviewmaketitle

\begin{figure*}[!tb] 
    \centering
    \subcaptionbox{1D imaging model for a planar sensor with a coded mask placed at distance $d$. Light rays from a light source at location $(\theta,z)$ are received by all the sensor pixels. A light ray that hits sensor pixel $s$ passes through mask at location $m$.}{  
    \includegraphics[width=.2\linewidth,keepaspectratio]{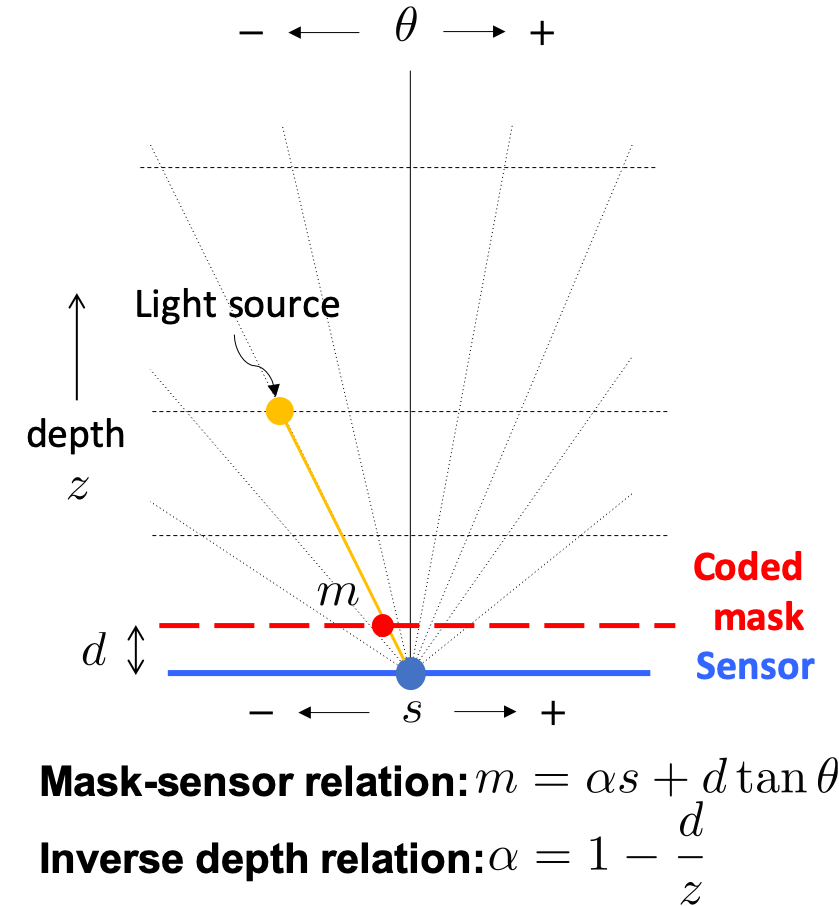}}
    \hspace{5pt} 
    \subcaptionbox{An overview of the proposed intensity and depth estimation framework. Consider a natural scene as a 3D point cloud, where each point represents a light source located at a different depth. The camera consists of a fixed, coded mask placed on top of an image sensor. Every point in the scene casts a shadow of the mask on the sensor plane. Each sensor pixels records a linear combination of the scene modulated by the mask pattern. The recovery algorithm consists of two steps. (1) Initialization using a greedy depth selection method. (2) An alternating gradient descent-based refinement algorithm that jointly estimates the light distribution and depth map on a continuous domain.}{    
    \includegraphics[width=0.7\linewidth, keepaspectratio]{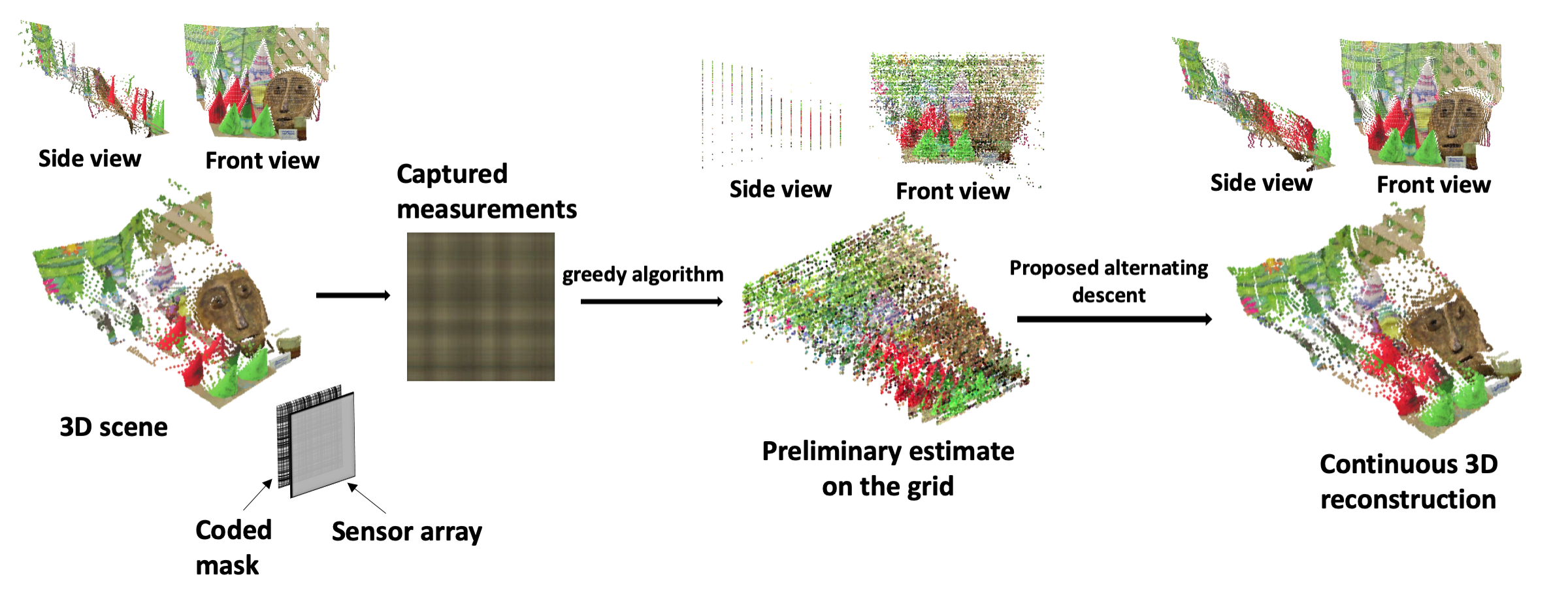}}
    \caption{A coded mask-based imaging model and an overview of the proposed continuous depth estimation framework.}\label{fig:intro}
\end{figure*}

\section{Introduction}

Depth estimation is an important and challenging problem that arises in a variety of applications including computer vision, robotics, and autonomous systems. Existing depth estimation systems use  stereo pairs of conventional (lens-based) cameras or time-of-flight sensors \cite{hartley2003multiple,gokturk2004tof,heide2013low}. These cameras can be heavy, bulky, and require large space for their installation. Therefore, their adoption in portable and lightweight devices with strict physical constraints is still limited. 

In this paper, we propose a joint image and depth estimation framework for a computational lensless camera that consists of a fixed, binary mask placed on top of a bare sensor. 
Such mask-based cameras offer an alternative design for building cameras without lenses. A recent example of a mask-based lensless camera is known as FlatCam \cite{asif2017flatcam}. In contrast with a lens-based camera that is designed to map every point in the scene to a single pixel on the sensor, every sensor pixel in a FlatCam records light from many points in the scene. A single point source in the scene casts a shadow of the mask on the sensor, which shifts if the point moves parallel to the sensor plane and expands/shrinks if the point source moves toward/away from the sensor plane. The measurements recorded on the sensor thus represent a superposition of shifted and scaled versions of the mask shadows corresponding to light sources in different directions and depths. Image and depth information about the scene is thus encoded in the measurements, and we can solve an inverse problem to estimate both of them. 


To jointly estimate the depth and light distribution, we propose a two-step approach that consists of an initialization step and an alternating gradient descent step to minimize our objective. To preserve sharp edges in the image intensity and depth map, we include an adaptive regularization penalty in our objective function. 

An overview of the reconstruction framework is illustrated in Figure~\ref{fig:intro}. In this paper, we use the same sampling framework proposed in \cite{asif2017lensless3D}. We initialize the estimate of the depth map by selecting a single plane or solving the greedy algorithm proposed in \cite{asif2017lensless3D}. The greedy method assumes that the scene consists of a small number of depth planes and fails to recover scene with continuous depth variations. The method proposed in this paper can estimate continuous depth by minimizing an objective function with respect to image intensity and depth via alternating gradient descent. We present extensive simulation and real experimental results with different objects.

The main contributions of this paper are as follows. 
\begin{itemize}[topsep=2pt,itemsep=2pt,leftmargin=*]
  \setlength{\itemsep}{0pt plus 1pt}
    \item We propose a new computational framework for joint estimation of light intensity and depth maps from a single image of a mask-based lensless camera. In contrast to other methods, our method estimates the depth map on a continuous domain. Our algorithm consists of a careful initialization step based on greedy pursuit and an alternating minimization step based on gradient descent. 
    \item The problem of joint image and depth recovery is highly nonconvex. To tackle this issue, we present different regularization schemes that offer robust recovery on a diverse dataset.  
    \item We present simulation results on standard 3D datasets and demonstrated a significant improvement over existing methods for 3D imaging using coded mask-based lensless cameras. 

    \item We built a hardware prototype to capture measurements of real objects. We present image and depth reconstruction results of these real objects using our proposed algorithm and a comparison with existing methods.
\end{itemize}

\section{Related Work}
 
A pinhole camera, also known as \textit{camera obscura}, is the simplest example of a mask-based lensless camera. Even though a pinhole can easily provide an image of the scene onto a sensor plane, the image quality is often severely affected by noise because the amount of light collected is limited by the pinhole aperture \cite{yedidia2018analysis_aperture}. Coded aperture-based lensless cameras avoid this problem by increasing the number of pinholes and allowing more light to reach the sensor \cite{fenimore1978ura, busboom1998ura, cannon1980coded_aperture,asif2017flatcam, boominathan2016lensless,antipa2018diffusercam}. In contrast to a pinhole camera where only one inverted image of the scene is obtained through a single pinhole, the measurements captured through a coded-mask are a linear combination of all the pinhole images under every mask element. To recover an image of the scene, we need to solve a computational image recovery problem \cite{fenimore1978ura, asif2017flatcam, antipa2018diffusercam}. 

Recent work on mask-based lensless imaging broadly falls into two categories. FlatCam \cite{asif2017lensless3D} uses a separable mask aligned with the sensor such that the sensor measurements corresponding to a plane at a fixed depth from the sensor can be written as a separable system. DiffuserCam \cite{antipa2018diffusercam} assumes that the mask size and angular span of the object are small enough so that the sensor measurements of a plane can be modeled as a convolution of the mask pattern with image intensity at that plane. The convolutional model can be computationally efficient if the object falls within a small angular range because we can use fast Fourier transform to compute convolutions. The separable model does not require a small angular range assumption. 
A number of methods based on deep learning have also been developed recently for both separable and convolutional imaging models to recover images at a fixed depth plane \cite{khan2019iccv, kris2019learned, dave2019deeppixel}. 

A coded aperture system offers another advantage by encoding light from different directions and depths differently.
The depth-dependent imaging capability in coded aperture systems is known since the pioneering work in this domain~\cite{barrett1973fresnelzoneImaging,fenimore1978ura}. However, the classical methods usually assume that the scene consists of a single plane at known depth. In this paper, we assume that the depth map is arbitrarily distributed on a continuous domain and the true depth map is unknown at the time of reconstruction.

The 3D lensless imaging problem has also recently been studied in \cite{antipa2018diffusercam,asif2017lensless3D,adams2017rice_depth,boominathan2016lensless,yihua2020sweep}. 
These methods can broadly be divided into two categories.  In the first category, the 3D scene is divided into a finite number of voxels. To recover the 3D light distribution, these methods solve an $\ell_1$ norm-based recovery problem under the assumption that the scene is very sparse  \cite{antipa2018diffusercam,adams2017rice_depth}. 
In the second category, the 3D scene is divided into an intensity map and multiple depth planes such that each pixel is assigned one intensity and depth. To solve the intensity and depth recovery problem, these methods either sweep through the depth planes \cite{boominathan2016lensless, yihua2020sweep} or assign depth to each pixel using a greedy method \cite{asif2017lensless3D}. Our proposed method belongs to the second category in which we model the image intensity and depth separately and assume that the depth values of the scene are distributed on a continuous domain. To recover the 3D scene, we jointly estimate the image intensity and depth map from the available sensor measurements. 

Joint estimation of image intensity and depth map can be viewed as a nonlinear inverse problem in which the sampling function is dependent on scene depth. Similar inverse problem also arises in many other fields such as direction-of-arrival estimation in radar \cite{tan2014off_grid_signal}, super-resolution \cite{boyd2015super_resolution} and compressed sensing \cite{tropp2007omp, needell2010cosamp, baraniuk2010model}. Similar to the joint estimation of image intensity and depth, the solution approaches to these problems consists of two main steps: identification of signal bases and the estimation of signal intensities based on the identified bases. The problem of identifying the signal bases from continuously varying candidates is often called off-the-grid signal recovery. The methods for solving the off-the-grid signal recovery problems can be divided into two main types. The first approach formulates the problem as a convex program on a continuous domain and solves it using an atomic norm minimization approach \cite{chandrasekaran2012atomic_norm, tang2013off_grid_cs}. The second approach linearizes the problem for the optimization parameter using a first-order approximation at every iteration \cite{boyd2015super_resolution, yang2013off_grid_doa}. Our proposed algorithm is inspired by the second approach.

Mask-based lensless cameras have traditionally been used for imaging light at wavelengths beyond the visible spectrum \cite{busboom1998ura, cannon1980coded_aperture}. 
Other examples related to mask-based cameras include controllable aperture and employing coded-mask for compressed sensing and computational imaging \cite{takhar2006single_pixel, zomet2006controllable_aperture}, distributed lensless camera\cite{yucheng2019asilomar}, single pixel camera \cite{duarte2008single_pixel} and external mask setting \cite{reddy2013external}. 

Coded masks have also recently been used with conventional lens-based cameras to estimate depth and lightfield \cite{levin2007_mask_on_lens,veeraraghavan2007dappled,marwah2013lightfield, hirsch2014lightfield}. Recently, a number of data-driven methods have been proposed to design custom phase masks and optical elements to estimate depth from a single image \cite{wu2019phasecam, chang2019deepOptics3D}. An all-optical diffractive deep neural network is proposed in \cite{lin2018optical_learning, mengu2020diffractive}, which can perform pattern recognition tasks such as handwritten digits classification using optical mask layers. Such networks can literally process images at a lightning-fast pace with near-zero energy cost.

\section{Methods}
\subsection{Imaging Model}\label{sec:model}
We divide the 3D scene under observation into $N\times N$ uniformly spaced directions.
We use $\theta_i$ and $\theta_j$ to denote the angular directions of a light source with respect to the center of the sensor. The intensity and depth of the light source are denoted using $l_{i,j}$ and $z_{i,j}$ respectively. 
Figure~\ref{fig:intro}(a) depicts the geometry of such an imaging model. A planar coded-mask is placed on top of a planar sensor array at distance $d$. The $M\times M$ sensor array captures lights coming from the scene modulated by the coded-mask. 

Every light source in the scene casts a shadow of the mask on the sensor array, which we denote using basis functions $\psi$. We use $s_u$ and $s_v$ to index a pixel on the rectangular sensor array. The shadow cast by a light source with unit intensity at $(\theta_i,\theta_j,z_{i,j})$ can be represented as the following basis or point spread function:  
\begin{equation}\label{eq:maskFunction}
    \psi_{i,j}(s_u,s_v)= \text{mask}\ [\alpha_{i,j} s_u+d\tan(\theta_i),\alpha_{i,j} s_v+d\tan(\theta_j)], 
\end{equation}
where $\text{mask}[u,v]$ denotes the transmittance of the mask pattern at location $(u,v)$ on the mask plane and $\alpha_{i,j}$ is a variable that is related to the physical depth $z_{i,j}$ with the following inverse relation: \begin{equation}
    \alpha_{i,j}=1-\frac{d}{z_{i,j}},
    \label{eq:depth_relation}
\end{equation}
If the 3D scene consists of only a single point source at $(\theta_i, \theta_j)$ with light intensity $l_{i,j}$, the measurement captured at sensor pixel $(s_u,s_v)$ would be
\begin{equation}\label{eq:single_point_source}
    y(s_u,s_v) =\psi_{i,j}(s_u,s_v)l_{i,j}.
\end{equation}

The measurement recorded on any sensor pixel is the summation of contributions from each of the point sources in the 3D scene. The imaging model for a single sensor pixel can be represented by 
\begin{equation}\label{eq:imagingModel} 
    y(s_u,s_v)=\sum_{i=1}^{N}\sum_{j=1}^{N}\psi_{i,j}(s_u,s_v)l_{i,j}.
\end{equation}
We can write the imaging model for the entire sensor in a compact form as 
\begin{equation}
    \mathbf{y}=\mathbf{\Psi}(\boldsymbol{\alpha})\mathbf{l}+e,
    \label{eq:compact_imaging_model}
\end{equation}
where $\mathbf{y}\in \mathbb{R}^{M^2}$ is a vectorized form of an $M\times M$ matrix that denotes sensor measurements, $\mathbf{l}\in\mathbb{R}^{N^2}$ is a vectorized form of an $N\times N$ matrix that denotes light intensity from all the locations $(\theta_i,\theta_j, \alpha_{i,j})$, and $\mathbf{\Psi}$ is a matrix with all the basis functions corresponding to $\theta_i,\theta_j, \alpha_{i,j}$. 
The basis functions in \eqref{eq:compact_imaging_model} are parameterized by the unknown $\boldsymbol{\alpha} \in \mathbb{R}^{N^2}$ and $e$ denotes noise and other nonidealities in the system. 

We can jointly estimate light distribution $(\mathbf{l})$ and inverse depth map $(\balpha)$\footnote{$\balpha$ has an inverse relation with the depth map \eqref{eq:depth_relation}; therefore we refer to it as inverse depth map throughout the paper.} 
using the following optimization problem: 
\begin{equation}
    \underset{\balpha,\mathbf{l}}{\text{minimize}}\ \frac{1}{2}\|\mathbf{y}- \mathbf{\Psi}(\balpha) \mathbf{l}\|_2^2.
    \label{eq:final_optimization}
\end{equation}
Note that if we know the true values of $\balpha$ (or we fix it to something), then the problem in \eqref{eq:final_optimization} reduces to a linear least-squares problem that can be efficiently solved via standard solvers. On the other hand, if we fix the value of $\mathbf{l}$, the problem remains nonlinear with respect to $\balpha$. In the next few sections we discuss our approach for solving the problem in \eqref{eq:final_optimization} via alternating minimization. 

\subsection{Initialization}
Since the minimization problem in \eqref{eq:final_optimization} is not convex, a proper initialization is often needed to ensure convergence to a local minimum close to the optimal point. 
A na\"ive approach is to initialize all the point sources in the scene at the same depth plane. To select an initial depth plane, we sweep through a set of candidate depth planes and perform image reconstruction on one depth plane at a time by solving the following linear least squares problem: 
\begin{equation}
    \underset{\mathbf{l}}{\text{minimize}}\ \frac{1}{2}\|\mathbf{y}- \mathbf{\Psi}(\balpha) \mathbf{l}\|_2^2.
    \label{eq:light_linear_least squares}
\end{equation}
We evaluate the loss value for all the candidate depth planes and picked the one with the smallest loss as our initialized depth. 
The mask basis function in \eqref{eq:maskFunction} changes as we change $\balpha$, which has an inverse relation with the scene depth. We select candidate depth corresponding to uniformly sampled values of $\balpha$, which yields non-uniform sampling of the physical scene depth.
The single-depth initialization approach is computationally simple and provides a reasonable initialization of light distribution to start with, especially when the scene is far from the sensor.

Our second approach for initialization is the greedy method proposed in \cite{asif2017lensless3D}. 
Greedy algorithms are widely used for sparse signal recovery \cite{tropp2007omp, needell2010cosamp,baraniuk2010model}. Based on these algorithms, \cite{asif2017lensless3D} proposed a greedy depth pursuit algorithm for depth estimation from FlatCam \cite{asif2017flatcam}. The algorithm works by iteratively updating the depth surface that matches the observed measurements the best.

The depth pursuit method assumes that the scene consists of a small number of predefined depth planes. We start the program by initializing all the pixels at a single depth plane and the estimation of light intensities $\mathbf{l}$ based on the initialized depth map. The first step is to select new candidate values for $\balpha$. The new candidates are selected using the basis vectors that are mostly correlated with the current residual of the estimate. In the second step, new candidates for $\balpha$ are appended to the current estimate. We solve a least squares problem using the appended $\balpha$. In the third step, we prune the $\balpha$ by selecting $\alpha_{i,j}$ as the value corresponding to the largest magnitude of $\mathbf{l}_{i,j}$. Although this method may not estimate the off-grid point sources well, it produces a good preliminary estimate of the scene.

\subsection{Refinement via Alternating Gradient Descent}\label{sec:gradients}
To solve the minimization problem in \eqref{eq:final_optimization}, we start with the preliminary image and depth estimates from the initialization step and alternately update depth and light distribution via gradient descent.
The main computational task in gradient descent method is computing the gradient of the loss function w.r.t. $\balpha$. To compute that gradient, we expand the loss function in \eqref{eq:final_optimization} as 
\begin{align}\label{eq:loss_scalars}
L &= \frac{1}{2}\sum_{u,v=1}^{M}(y(s_u,s_v)- \sum_{i,j=1}^{N}\psi_{i,j}(s_u,s_v)l_{i,j})^2 
\end{align}
We define $R_{u,v}=y(s_u,s_v)-\sum_{i,j=1}^{N}\psi_{i,j}(s_u,s_v)l_{i,j}$ as the residual approximation error at location $(s_u,s_v)$. The derivatives of the loss function with respect to the $\alpha_{i,j}$ is given as
\begin{align}\label{eq:L_derive}
        \frac{\partial L}{\partial \alpha_{i,j}}&=
        \sum_{u,v=1}^{M} R_{u,v}\frac{\partial R_{u,v}}{\partial \alpha_{i,j}}
        =-l_{i,j}\sum_{u,v=1}^{M} R_{u,v}\frac{\partial \psi_{i,j}(s_u,s_v)}{\partial \alpha_{i,j}}. 
\end{align}

We compute the derivatives of sensor value with respect to the $\alpha_{i,j}$ using the total derivative \footnote{Recall that the total derivative of a multivariate function $f(x,y)$ is $\frac{\partial f(x,y)}{\partial x}dx+\frac{\partial f(x,y)}{\partial y}dy$.} as follows. 
\begin{align}\label{eq:gradient_derivations}
    \frac{\partial \psi_{i,j}(s_u,s_v)}{\partial \alpha_{i,j}} 
    &= \frac{\partial \psi_{i,j}(s_u,s_v)}{\partial u_{i,j}}\frac{\partial u_{i,j}}{\partial\alpha_{i,j}}
    +\frac{\partial \psi_{i,j}(s_u,s_v)}{\partial v_{i,j}}\frac{\partial v_{i,j}}{\partial\alpha_{i,j}} \nonumber \\
    & =
    \frac{\partial \psi_{i,j}(s_u,s_v)}{\partial u_{i,j}}s_u 
    +\frac{\partial \psi_{i,j}(s_u,s_v)}{\partial  v_{i,j}}s_v.
\end{align}
$u_{i,j} = \alpha_{i,j} s_u+d\tan(\theta_i)$ and $v_{i,j} = \alpha_{i,j} s_v+d\tan(\theta_j)$ denote two dummy variables that also correspond to the specific location on the mask where a light ray from a point source at angle $(\theta_i,\theta_j)$ and depth $\alpha_{i,j}$ and sensor pixel at $(s_u,s_v)$ intersects with the mask plane. 
The terms in $\frac{\partial \psi_{i,j}(s_u,s_v)}{\partial u_{i,j}}, \frac{\partial \psi_{i,j}(s_u,s_v)}{\partial v_{i,j}}$ can be viewed as the derivatives of mask pattern along the respective spatial coordinates and evaluated at $u_{i,j},v_{i,j}$. We compute these derivatives using finite-difference of $\psi_{i,j}(s_u,s_v)$ over a fine grid and linear interpolation. 

\subsection{Algorithm Analysis}\label{sec:algorithmanalysis}
To solve the non-linear least squares problem in \eqref{eq:final_optimization} in our algorithms, we compute the gradient derived in \eqref{eq:gradient_derivations} and use it as input of a optimization solver. Suppose $\psi_i$ and $\psi_j$ denote the basis function vectors evaluated on a 1D mask as
\begin{align}
    \psi_i(s_u) &= \text{mask}\ [\alpha_{i,j} s_u+d\tan(\theta_i)] \nonumber \\
    \psi_j(s_v) &= \text{mask}\ [\alpha_{i,j} s_v+d\tan(\theta_j)]. 
\end{align}
If we use a separable mask pattern, then the 2D mask function $\psi_{i,j}$ in \eqref{eq:maskFunction} can be computed as the outer product of two vectors given as $\psi_{i,j}=\psi_i\psi_j^T$. Similarly, we define 1D sub-gradient function $g$ as
\begin{align}
    g_i(s_u) &=  \frac{\partial \psi_{i,j}(s_u,s_v)}{\partial u_{i,j}} \nonumber \\
    g_j(s_v) &= \frac{\partial \psi_{i,j}(s_u,s_v)}{\partial  v_{i,j}}, 
\end{align}
Similar to \eqref{eq:gradient_derivations}, the functions $\frac{\partial \psi_{i,j}(s_u,s_v)}{\partial u_{i,j}}$ and $\frac{\partial \psi_{i,j}(s_u,s_v)}{\partial  v_{i,j}}$ are the sub-gradient functions along the 1D mask. It takes non-negative values at locations where mask pattern value changes and takes zero value at the other places. Using the derivation in \eqref{eq:gradient_derivations}, the matrix contains $\frac{\partial \psi_{i,j}(s_u,s_v)}{\partial \alpha_{i,j}}$ at all $(s_u,s_v)$ can be computed using the following sum of two vector outer products.
\begin{equation}
    \frac{\partial \psi_{i,j}}{\partial \alpha_{i,j}} = g_i\psi_j^T+\psi_i g_j^T
\end{equation}
Using the derivations in \eqref{eq:L_derive}, the derivative of loss function with respect to depth value can be computed using the following matrix multiplications, where $R$ refers to the matrix of residual $R_{u,v}$ at all $(s_u,s_v)$
\begin{equation}
    \frac{\partial L}{\partial \alpha_{i,j}}=g_i^T R\psi_j + \psi_i^T R g_j
    \label{eq:vector_gradient}
\end{equation}
Suppose we have $M\times M$ pixels on sensor array. The computation in \eqref{eq:vector_gradient} takes $2M^2+2M$ multiplications.
We then feed our gradients to \texttt{minfunc} solver \cite{schmidt2005minfunc} with L-BFGS algorithm \cite{liu1989lbfgs} to solve the non-linear optimization problem in \eqref{eq:final_optimization}.

\subsection{Regularization Approaches}
\begin{figure}[t]
    \centering
    \begin{subfigure}{0.45\linewidth}
    \includegraphics[width=1\linewidth,keepaspectratio]{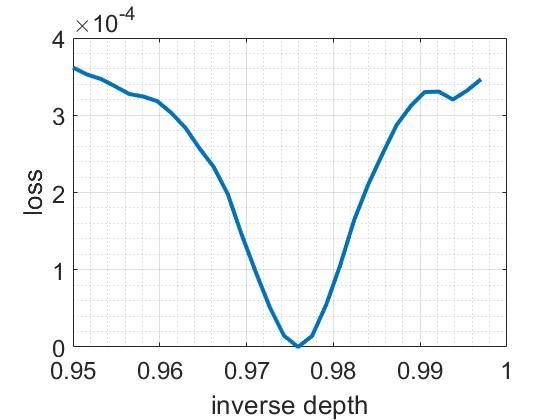}
    \caption{Without smooth regularization, the loss curve is highly non-convex and contains several local minima.}
    \end{subfigure}
    \begin{subfigure}{0.45\linewidth}
    \includegraphics[width=1\linewidth,keepaspectratio]{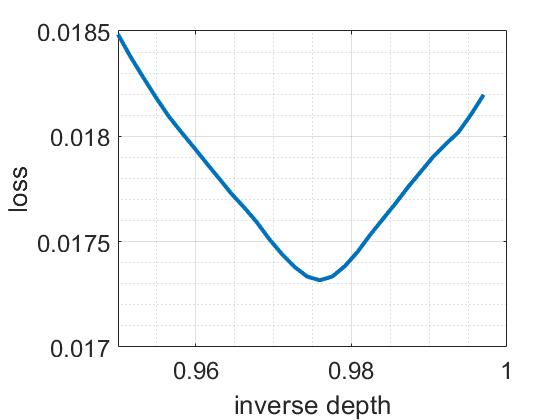}
    \caption{With the smooth regularization, the loss curve is smooth and several local minima are removed.} 
    \end{subfigure}
    \begin{subfigure}{0.45\linewidth}
    \includegraphics[width=1\linewidth,keepaspectratio]{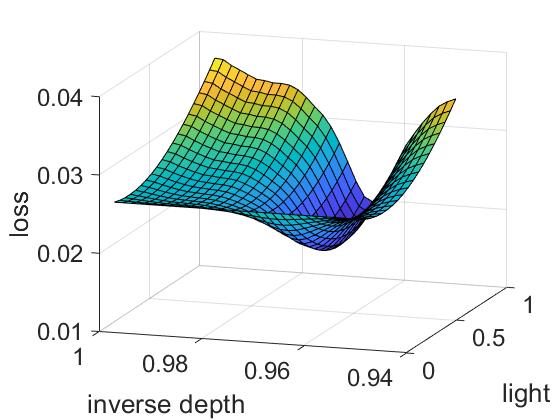}
    \caption{Similar to 1D case, loss surface contains many local minima without smooth regularization.} 
    \end{subfigure}
    \begin{subfigure}{0.45\linewidth}
    \includegraphics[width=1\linewidth,keepaspectratio]{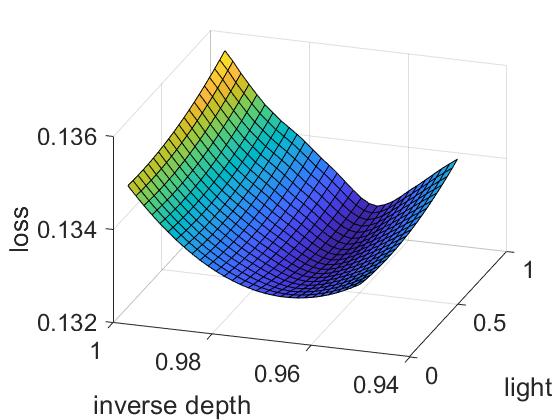}
    \caption{With smooth regularization, many local minima are removed from loss surface. } 
    \end{subfigure}
    \caption{A comparison between objective loss functions without and with smooth regularization. The inverse depth axis refers to the value of $\alpha$.}
    \label{fig:regularization}
\end{figure}

\noindent{\bf $\ell_2$ regularization on spatial gradients.} The optimization problem in \eqref{eq:final_optimization} is highly non-convex and contains several local minima; therefore, the estimate often gets stuck in some local minima and the estimated intensity and depth maps are coarse. 
To improve the performance of our algorithm for solving the non-convex problem in  \eqref{eq:final_optimization}, we seek to exploit additional structures in the scene. A standard assumption is that the depth of neighboring pixels is usually close, which implies that the spatial differences of (inverse) depth map are small. To incorporate this assumption in our model, we add a quadratic regularization term on the spatial gradients of the inverse depth map to our loss function. 
The quadratic regularization term is defined on an $N\times N$ inverse depth map matrix $\balpha$ and can be written as 
\begin{align}
        R(\balpha) &=\sum_{i,j=1}^{N} (\alpha_{i,j}-\alpha_{i+1,j})^2+(\alpha_{i,j}-\alpha_{i,j+1})^2 \nonumber \\
        &=\|\nabla_r\balpha\|_F^2+\|\nabla_c\balpha\|_F^2,
        \label{eq:quadratic_reg}
\end{align}
where the operators $\nabla_r, \nabla_c$ compute spatial differences along rows and columns, respectively. We call this regularization an $\ell_2$ norm-based total variation (TV-$\ell_2$) in this paper. 
Figure~\ref{fig:regularization} illustrates the effect of the depth  regularization. From Figure~\ref{fig:regularization}, we observe that smoothness regularization improves the loss function by removing several local minima. We also observed this effect in our simulations for a high-dimensional depth recovery problem, which is not very sensitive to initialization with depth regularization. 

\noindent{\bf Weighted $\ell_2$ regularization on spatial gradients.}
Even though smoothness regularization on the inverse depth map removes some local minima and helps with converge, it does not respect the sharp edges in the depth map. To preserve sharp discontinuities in the (inverse) depth map, we used the following adaptive weighted regularization inspired from \cite{liu2012weighted_smooth}: 
\begin{equation}
    R_W(\balpha) = \sum_{i,j=1}^{N} W_{i,j}^{c}(\alpha_{i,j}-\alpha_{i+1,j})^2+W_{i,j}^{r}(\alpha_{i,j}-\alpha_{i,j+1})^2, 
    \label{eq:weighted_depth_smooth_reg}
\end{equation}
where $W^{r,\alpha}_{i,j}$ and $W^{c,\alpha}_{i,j}$ denote weights for row and column differences, respectively. We aim to select these weights to promote depth similarity for neighboring pixels, but avoid smoothing the sharp edges. 
To promote this, we selected weights with exponential decay in our experiments that we compute as 
\begin{align}
    W_{i,j}^{r} &= \exp\left(-\frac{(\alpha_{i,j}-\alpha_{i+1,j})^2}{\sigma}\right) \nonumber \\
    W_{i,j}^{c} &= \exp\left(-\frac{(\alpha_{i,j}-\alpha_{i,j+1})^2}{\sigma}\right).
    \label{eq:weight_computation_depth}
\end{align}
Such a weighted regularization forces pixels that have depth within a small range of one another to be smooth and does not penalize the points that have larger gap in depth (which indicates the presence of an edge). This helps preserve sharp edges in the reconstructed depth estimates. This weighting approach is analogous to bilateral filtering approach for image denoising \cite{tomasi1998bilateral, durand2002bilateral}.

The regularized estimation problem for image and depth can be written in the following form: 
\begin{equation}
    \underset{\balpha,\mathbf{l}}{\text{minimize}}\; \frac{1}{2}\|\mathbf{y}- \mathbf{\Psi}(\balpha) \mathbf{l}\|_2^2 + \lambda R_W(\balpha).
    \label{eq:quadratic_reg_problem}
\end{equation}
We call this regularization approach weighted TV-$\ell_2$ and solve it by alternately updating the inverse depth map $\balpha$ and light intensity $\mathbf{l}$. A pseudocode of the algorithm is presented at Algorithm~\ref{alg:l2_algorithm}.

\renewcommand{\algorithmicrequire}{\textbf{Input:}}
\renewcommand{\algorithmicensure}{\textbf{Output:}}
\begin{algorithm}[htb] 
	\caption{Weighted TV-$\ell_2$ regularized optimization} 
	\label{alg:l2_algorithm}
	\begin{algorithmic}
		\REQUIRE{
		Sensor measurements: $\mathbf{y}$}
		\ENSURE {
		Light distribution and inverse depth map: $\mathbf{l}, \boldsymbol{\alpha}$}
		
		\textbf{Initialization via greedy algorithm:}
		\STATE Compute $\boldsymbol{\alpha}$ and $\mathbf{l}$ with depth pursuit algorithm in \cite{asif2017lensless3D}.
        
        \textbf{Refinement via alternating gradient descent:}
		\FOR {$k=1:k_{\text{max}}$}
                	\item $	\widehat{\boldsymbol{\alpha}}^k=\underset{\boldsymbol{\alpha} }{\text{argmin}}  \frac{1}{2}\|\mathbf{y}-\mathbf{\Psi}(\boldsymbol{\alpha})\mathbf{l}^{k-1}\|_2^2 + \lambda R_W(\boldsymbol{\alpha}) $

        \item $	\widehat{\mathbf{l}}^k =\underset{\mathbf{l}}{\text{argmin}} \ \frac{1}{2}\|\mathbf{y}-\mathbf{\Psi}(\boldsymbol{\alpha}^{k})\mathbf{l}\|_2^2$
		\ENDFOR
		\RETURN $\widehat{\mathbf{l}}$ and $\widehat{\boldsymbol{\balpha}}$
	\end{algorithmic}
\end{algorithm}

\noindent {\bf $\ell_1$ regularization on spatial gradients.} It is well-known that the $\ell_1$ norm regularization enforces the solution to be sparse. We add an $\ell_1$-based total variation norm \cite{rudin1992tv} of the depth to our optimization problem. By enforcing the sparsity of spatial gradients, the edges of (inverse) depth map can be preserved. The $\ell_1$ norm-based TV regularization term is given as
\begin{align}
    R_{TV}(\balpha) &=\sum_{i,j=1}^N |\alpha_{i,j}-\alpha_{i+1,j}|+|\alpha_{i,j}-\alpha_{i,j+1}| \nonumber \\
    & = \|\nabla_r \balpha \|_1 + \|\nabla_c \balpha\|_1. 
\end{align}
To solve the nonlinear optimization problem with $\ell_1$ norm regularization, we write the optimization problem as 
\begin{align}
            \underset{\balpha,\mathbf{l}}{\text{minimize}}\;  &\frac{1}{2}\|\mathbf{y}- \mathbf{\Psi}(\balpha) \mathbf{l}\|_2^2 + \lambda(\|\mathbf{d}_r\|_1+\|\mathbf{d}_c\|_1) \nonumber \\
            \text{s.t.}~  &\mathbf{d}_r=\nabla_r\balpha,~\mathbf{d}_c=\nabla_c\balpha.
            \label{eq:l1_reg}
\end{align}
We solve this problem \eqref{eq:l1_reg} using a split-Bregman method \cite{split-bregman}. 

\subsection{Computational complexity}
The main computational and storage cost of the proposed method arises from the forward and adjoint of the imaging operators. Since the depth of any pixel can have an arbitrary value in a continuous domain, we cannot precompute the imaging operators. At every iteration, we first interpolate the separable mask patterns according to the current estimate of $\alpha$. The time complexity of forward imaging operator as given in \eqref{eq:compact_imaging_model} will be  $O(M^2N^2)$ because we have to add up depth-dependent response corresponding to every angle.

\section{Simulation Results} \label{sec:simulations}
In this section, we present simulation results to evaluate the performance of our methods under different noise levels and sensor sizes. We also present a comparison of our proposed method with two existing methods for 3D imaging with lensless cameras. 
Additional experiments on the reconstruction of a single depth plane and the effect of numbers of sensor pixels on the reconstruction are included in the Supplementary material.

\subsection{Simulation Setup}
To validate the performance of the proposed algorithm, we simulate a lensless imaging system using a binary planar mask with a separable maximum length sequence (MLS) pattern \cite{macwilliams1976mls} that is placed 4mm away from a planar sensor array. 
We used an MLS sequence of length 1024 and converted all the $-1$s to 0s to create a separable binary pattern. We used square mask features, each of which is 30$\mu$m wide. Since we optimize the objective function in \eqref{eq:final_optimization} with respect to $\alpha$ and need to compute the gradient in \eqref{eq:L_derive}, we require the mask function to be smooth and differentiable with respect to $\alpha$. Therefore, we convolved the binary pattern with a Gaussian blur kernel of length 15$\mu$m and standard deviation 5. 
In our simulations, we do not explicitly model the diffraction blur. However, the Gaussian blur kernel that we apply to the mask function can be viewed as an approximation of the diffraction blur.
The sensor contains $512\times512$ square pixels, each of which is 50$\mu$m wide. The chief ray angle of each sensor pixel is $\pm \ang{18}$. We assume that there is no noise added to the sensor measurements. 
In our experiments for continuous depth estimation, we fixed all the parameters to these default values and analyze the performance with respect to a single parameter.

\begin{figure}[thb]
    \centering
    \begin{subfigure}{0.3\linewidth}
    \caption*{Original}
    \includegraphics[width=1\linewidth,keepaspectratio]{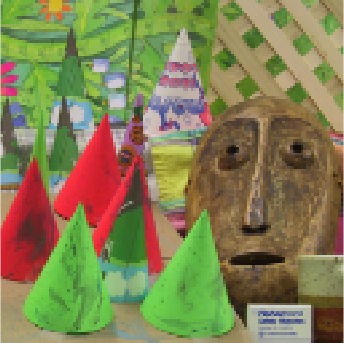}
    \caption*{Image PSNR:}
    \includegraphics[width=1\linewidth,keepaspectratio]{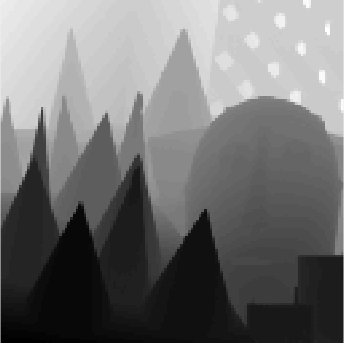}
        \caption*{Depth RMSE:}
    \end{subfigure}
    \begin{subfigure}{0.3\linewidth}
    \caption*{Greedy\cite{asif2017lensless3D}} 
    \includegraphics[width=1\linewidth,keepaspectratio]{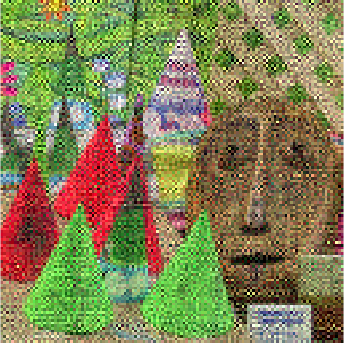}
    \caption*{16.57dB}
    \includegraphics[width=1\linewidth,keepaspectratio]{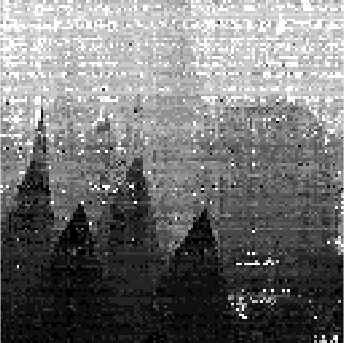}
        \caption*{87.48mm} 
    \end{subfigure}
        \begin{subfigure}{0.3\linewidth}
    	\caption*{Ours} 
    	\includegraphics[width=1\linewidth,keepaspectratio]{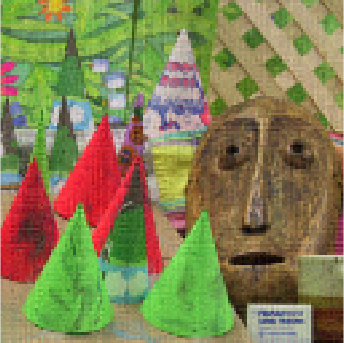}
    	\caption*{31.65dB}
    	\includegraphics[width=1\linewidth,keepaspectratio]{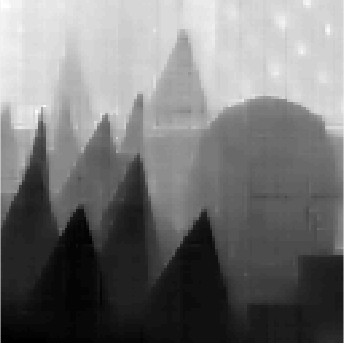}
    	\caption*{17.90mm} 
    \end{subfigure}
    
    \caption{Left to right: original image and depth of the Cones scene; image and depth initialized via greedy algorithm \cite{asif2017lensless3D}; depth estimation using weighted $\ell_2$-based regularization. The depth in this scene varies from around 0.99m to 1.7m. }
    \label{fig:cones_scene}
\end{figure}

\subsection{Reconstruction of Scenes with Continuous Depth}

\paragraph*{Depth datasets} We performed all our experiments on 3D images created using light intensities and depth information from Middlebury\cite{scharstein2001stereo}, Make3D\cite{saxena2006make3D_1,saxena2006make3D_2} and NYU Depth\cite{silberman2012NYU_depth}, the test scenes and their depth ranges are listed in Table~\ref{table:testing_scenes}.

\begin{table}[!h]
\centering
\begin{tabular}{ccc}
\textbf{Test datasets} & \textbf{Min depth (m)} & \textbf{Max depth (m)} \\ \hline
Sword                  & 0.65                  & 0.95                  \\
Playtable              & 1.47                  & 3.75                  \\
\hline 
Cones                  & 0.99                  & 1.70                  \\
Corner                 & 3.93                  & 10.60              \\
Whiteboard             & 1.08                  & 2.90                  \\
Playroom               & 1.62                  & 2.93                  \\
Moebius                & 0.74                  & 1.23                  \\
Books                   & 0.73                  & 1.27                  
\end{tabular}
\caption{Analysis experiments are performed on multiple scenes picked from Middlebury\cite{scharstein2001stereo}, Make3D\cite{saxena2006make3D_1,saxena2006make3D_2} and NYU Depth\cite{silberman2012NYU_depth}. Results of the two scenes above line are presented within the main text, while the rest of them are reported in the supplementary material. }
\label{table:testing_scenes}
\end{table}

\begin{figure*}
\captionsetup[subfigure]{position=b}
\centering
\begin{subfigure}[t]{0.23\linewidth}
    \caption*{Original Scene} 
    \includegraphics[width=1\linewidth, keepaspectratio]{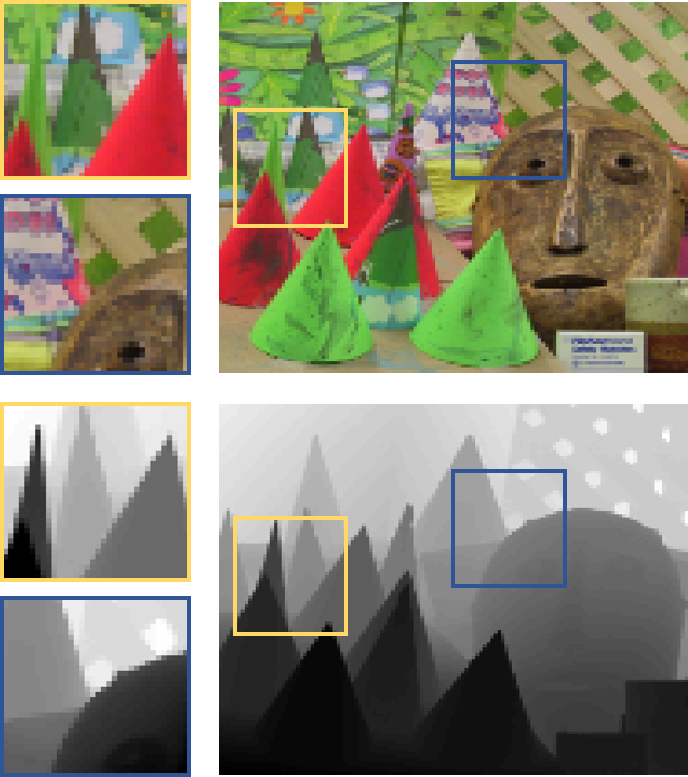}
    \caption{Image and depth of the original scene. The selected Cones scene is taken from Middlebury dataset \cite{scharstein2001stereo}. The range of depth is from 0.99 to 1.7 meters. }
\end{subfigure}
~
\begin{subfigure}[t]{0.23\linewidth}
    \caption*{TV-$\ell_2$} 
    \includegraphics[width=1\linewidth, keepaspectratio]{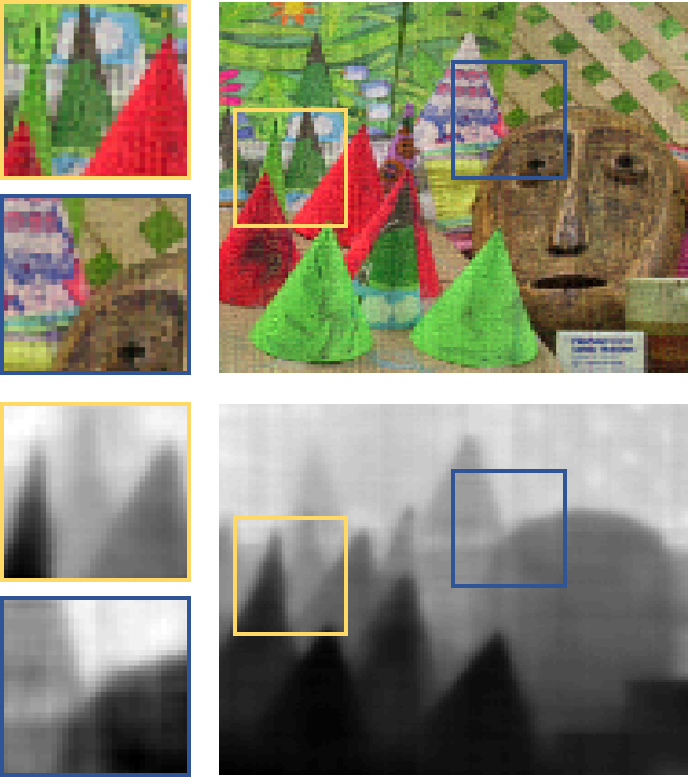}
    \caption{Image and depth reconstruction from isotripic total variation. PSNR of image is 29.69dB and depth RMSE is 25.21mm. }
\end{subfigure}
~
\begin{subfigure}[t]{0.23\linewidth}
    \caption*{Weighted TV-$\ell_2$} 
    \includegraphics[width=1\linewidth, keepaspectratio]{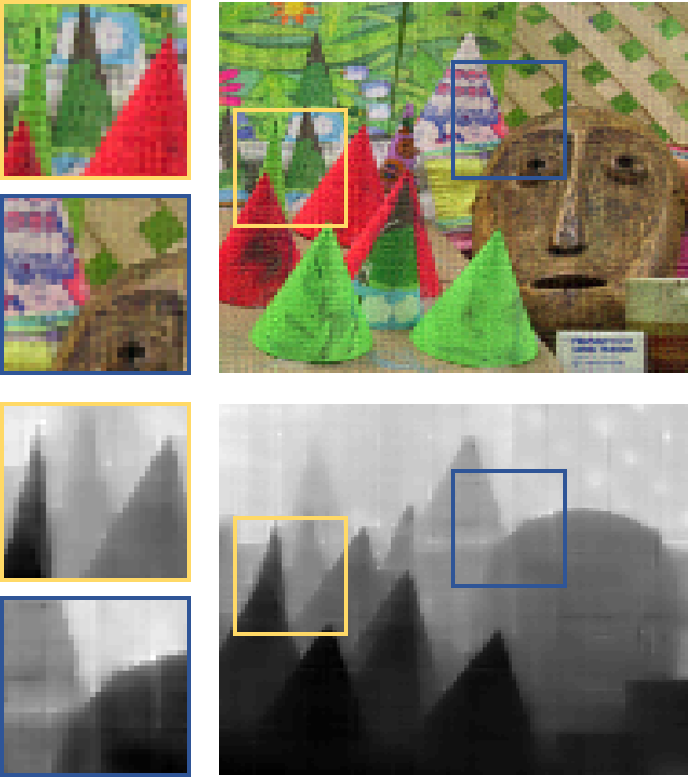}
    \caption{Image and depth reconstruction from weighted $\ell_2$ total variation. The PSNR of image is 31.65dB and the RMSE of depth is 17.90mm. The edges of depth are preserved better.}
\end{subfigure}
~~
\begin{subfigure}[t]{0.23\linewidth}
    \caption*{TV-$\ell_1$} 
    \includegraphics[width=1\linewidth, keepaspectratio]{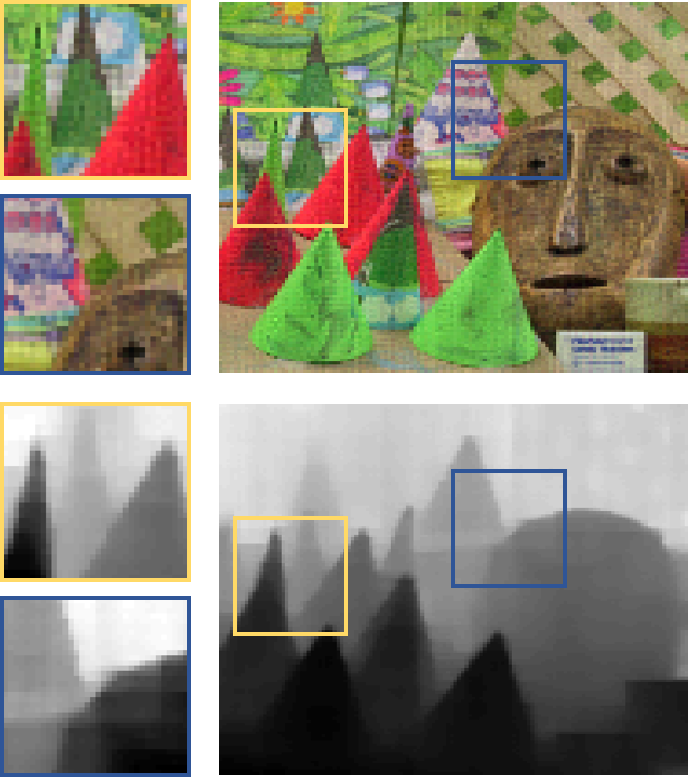}
    \caption{Image and depth reconstruction from TV-$\ell_1$. The PSNR of image is 30.82dB and the depth RMSE is 19.56mm. The edges of depth are preserved better.}
\end{subfigure}
\caption{
Comparison between reconstructions using three different regularization approaches from the same measurements.} 
\label{fig:reg_comparison}
\end{figure*}

\paragraph*{Initialization via greedy method} Let us further discuss our simulation setup using the \textit{Cones} scene, for which the results are presented in Figure~\ref{fig:cones_scene}. 
We simulated the 3D scene using depth data from Middlebury dataset\cite{scharstein2001stereo}. We sample the scene at uniform angles to create a $128\times128$ image and its (inverse) depth map with the same size. We can compute the physical depth from $\balpha$ using \eqref{eq:depth_relation}. In our simulation, the depth of this scene ranges from around 0.99m to 1.7m. We used depth pursuit greedy algorithm in \cite{asif2017lensless3D} as our initialization method. We selected 15 candidate depths by uniformly sampling the inverse depth values $\balpha$ from 0.996 to 0.9976, which gives an effective depth in the same range as the original depth. 
Since we are trying to gauge the performance for off-the-grid estimate of depth, the candidate values of $\balpha$ are not exactly the same as the true values of $\balpha$ in our simulations. The output of the initialization algorithm is then fed into the alternating gradient descent method. 

\paragraph*{Performance metrics} We evaluate the performance of recovered image intensity and depth independent of each other. We report the peak signal to noise ratio (PSNR) of the estimated intensity distribution and root mean squared error (RMSE) of the estimated depth maps for all our experiments. The estimates for image intensity and depth maps for the initialization and our proposed weighted TV-$\ell_2$ method are shown in Figure~\ref{fig:cones_scene}, along with the PSNR and RMSE. 
We can observe that both image and depth estimation from greedy method \cite{asif2017lensless3D} contain several spikes because of the model mismatch with the predefined depth grid. In contrast, many of these spikes are removed in the estimations from the proposed algorithm with weighted TV-$\ell_2$ while the edges are preserved.

\paragraph*{Comparison of regularization methods}
Here we present a comparison between three different regularization approaches. We reconstruct image intensity and (inverse) depth map using the same measurements with TV-$\ell_2$, weighted TV-$\ell_2$, and TV-$\ell_1$ regularization. The results are shown in Figure~\ref{fig:reg_comparison}. Compared to the TV-$\ell_2$ method, we observe that both weighted TV-$\ell_2$ and TV-$\ell_1$ preserve the sharp edges in image and depth estimates. Overall, in our experiments, weighted TV-$\ell_2$ provided the best results. Therefore, we used that as our default method for the rest of the paper.

\newcommand{\figwidth}{0.11\linewidth}
\begin{figure*}
    \centering
    \begin{subfigure}{\figwidth}
    \caption*{Original}
    \includegraphics[width=1\linewidth]{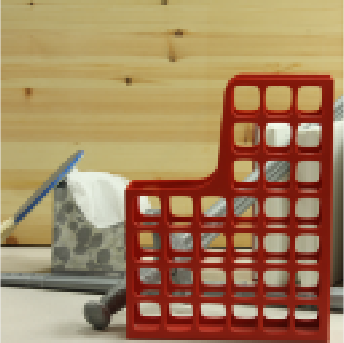}
    \caption*{PSNR:}
    \includegraphics[width=1\linewidth]{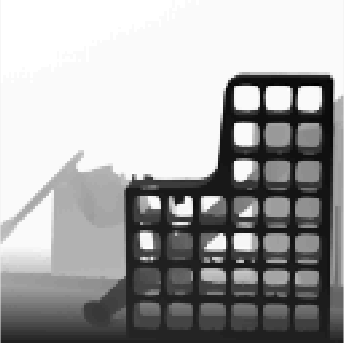}
        \caption*{RMSE:}
    \end{subfigure}
        \begin{subfigure}{\figwidth}
    \caption*{20dB} 
    \includegraphics[width=1\linewidth]{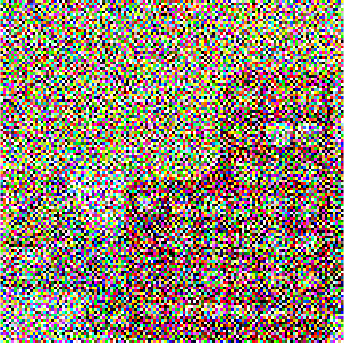}
            \caption*{1.15dB}
    \includegraphics[width=1\linewidth]{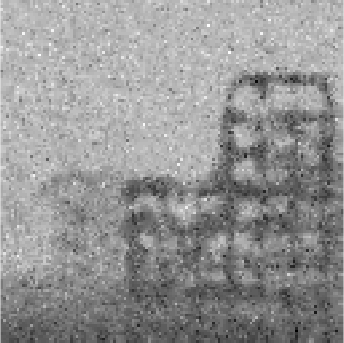}
        \caption*{66.98mm} 
    \end{subfigure}
        \begin{subfigure}{\figwidth}
    \caption*{30dB} 
    \includegraphics[width=1\linewidth]{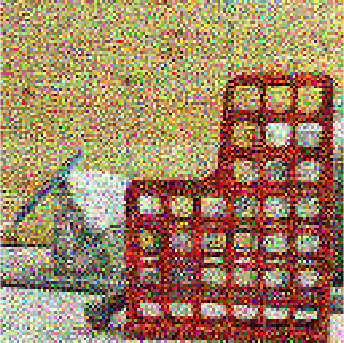}
     \caption*{11.33dB}
    \includegraphics[width=1\linewidth]{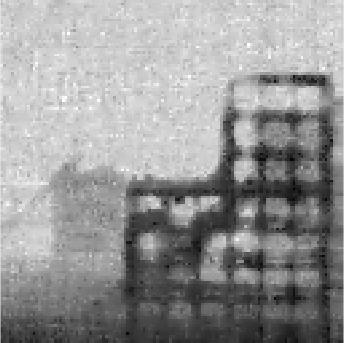}
        \caption*{38.84mm} 
    \end{subfigure}
        \begin{subfigure}{\figwidth}
    \caption*{40dB} 
    \includegraphics[width=1\linewidth]{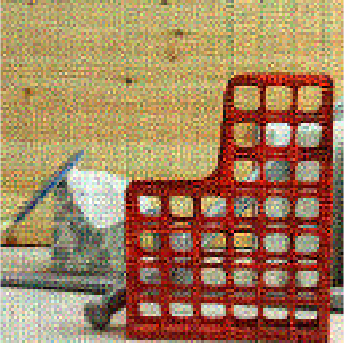}
     \caption*{19.16dB}
    \includegraphics[width=1\linewidth]{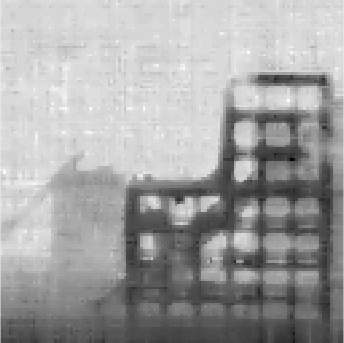}
        \caption*{34.29mm} 
    \end{subfigure}
   \begin{subfigure}{\figwidth}
       \caption*{Original}
    \includegraphics[width=1\linewidth]{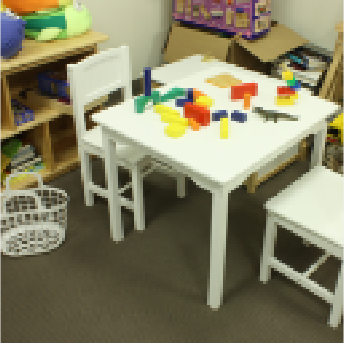}
     \caption*{PSNR:}
    \includegraphics[width=1\linewidth]{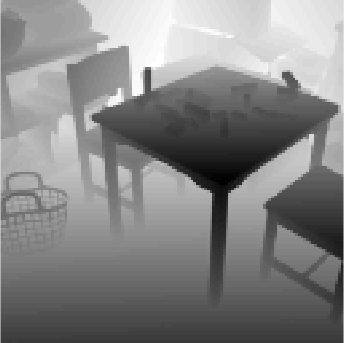}
        \caption*{RMSE:}
    \end{subfigure}
        \begin{subfigure}{\figwidth}
            \caption*{20dB} 
    \includegraphics[width=1\linewidth]{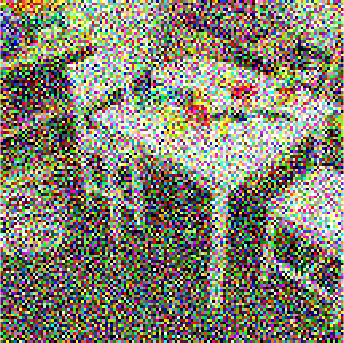}
    \caption*{5.12dB}
    \includegraphics[width=1\linewidth]{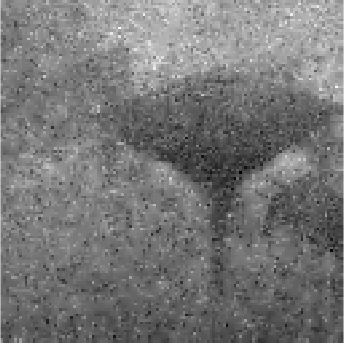}
        \caption*{360.95mm} 
    \end{subfigure}
        \begin{subfigure}{\figwidth}
        \caption*{30dB} 
    \includegraphics[width=1\linewidth]{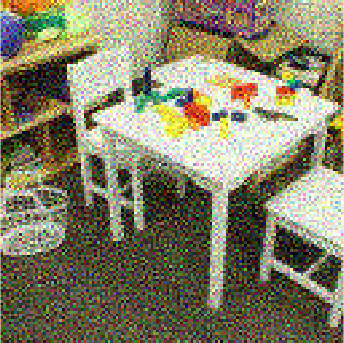}
    \caption*{15.50dB}
    \includegraphics[width=1\linewidth]{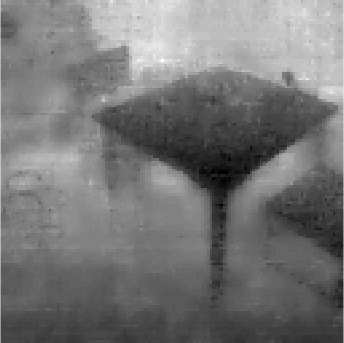}
        \caption*{163.26mm} 
    \end{subfigure}
        \begin{subfigure}{\figwidth}
        \caption*{40dB} 
    \includegraphics[width=1\linewidth]{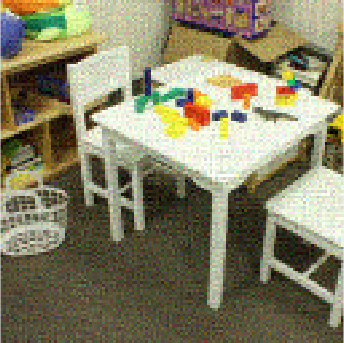}
    \caption*{24.88dB}
    \includegraphics[width=1\linewidth]{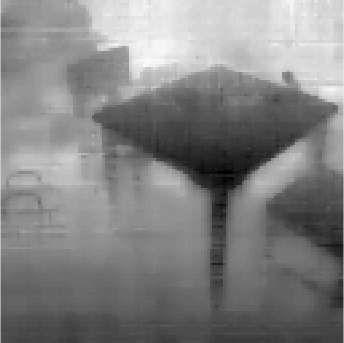}
        \caption*{151.46mm} 
    \end{subfigure}
    ~~~

    \caption{Effects of noise: Reconstruction from the measurements with signal-to-noise ratio (SNR) at 20dB, 30dB and 40dB, along with the \textbf{PSNR} of reconstructed image and \textbf{RMSE} of reconstructed depth map. As expected, the quality of reconstructed image and depth improves as the noise level is reduced. The sequence in left is for \textit{Sword}, right is \textit{Playtable}.}
    \label{fig:noise_examples}
\end{figure*}

\begin{figure}
    \centering
    \begin{subfigure}{0.49\linewidth}
    \includegraphics[width=1\linewidth]{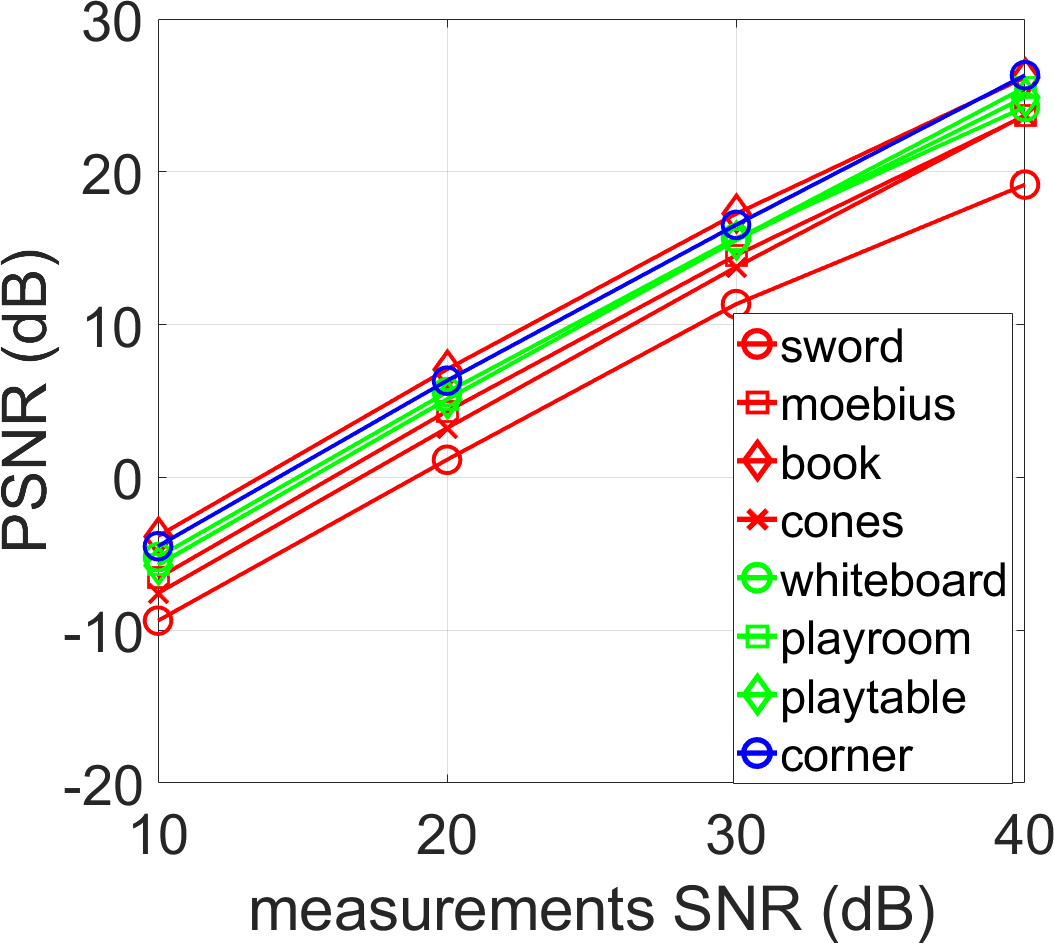}
    \caption{Image PSNR for different noise levels}
    \end{subfigure}
    \begin{subfigure}{0.49\linewidth}
    \includegraphics[width=1\linewidth]{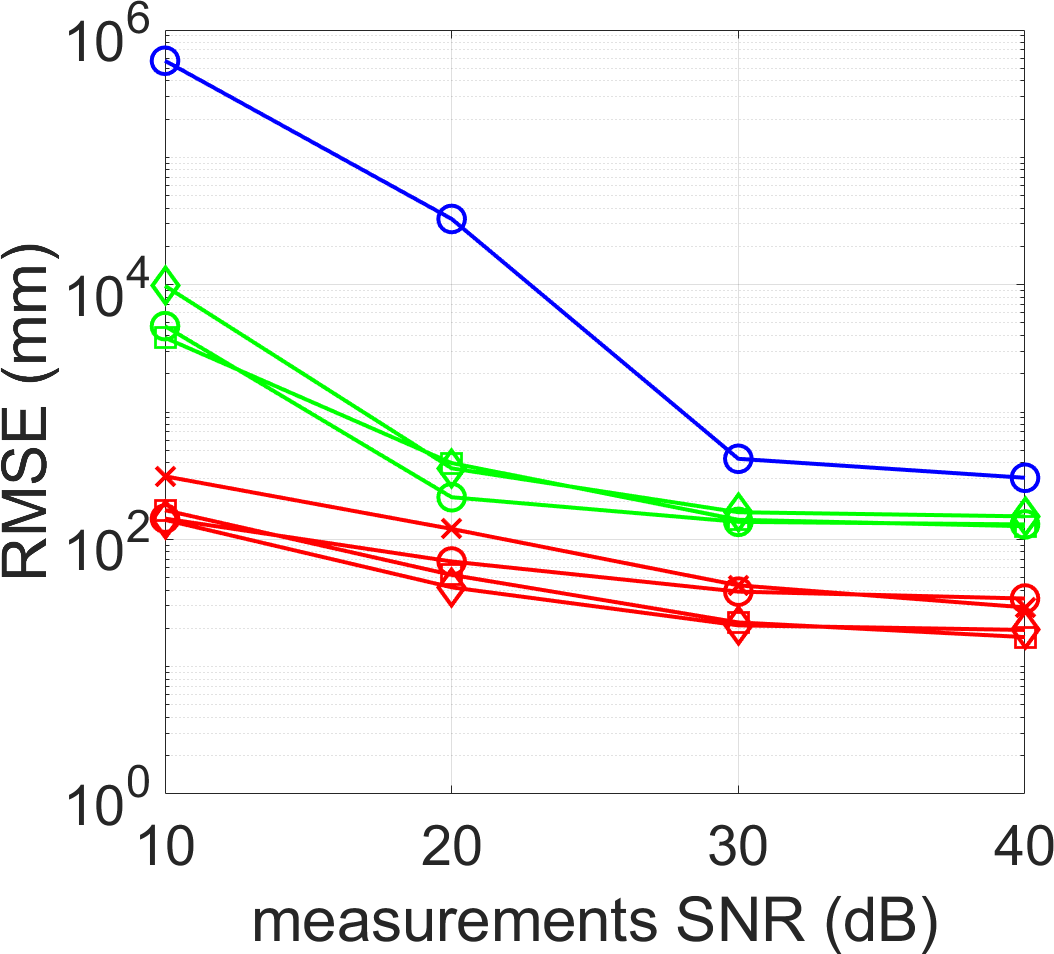}
    \caption{Depth RMSE for different noise levels}
    \end{subfigure}
    
    \caption{Reconstruction from measurements with different levels of Gaussian noise on multiple scenes. Both of the image Peak Signal-Noise Ratio and depth Root mean squared error are improved as the noise is reduced. The reconstruction quality degrades if the scene is placed farther from the camera.}
    \label{fig:noise_curve}
\end{figure}

\subsection{Effects of Noise}
Sensor noise exists widely in any observation process. The amplitude of noise depends on the intensities of sensor measurements and can adversely affect the reconstruction results.
To investigate the effect of noise on our algorithm,
we present simulation results for the reconstruction of scenes from the same sensor measurements under different levels of additive white Gaussian noise. The experiments are performed on multiple 3D scenes listed in Table~\ref{table:testing_scenes}. Some examples of reconstruction with different levels of noise are shown in Figure~\ref{fig:noise_examples}.

The plots recording PSNR of image intensities and RMSE of depth maps over a range of measurement SNR values are presented in Figure~\ref{fig:noise_curve}. As we can observe from the curves that the quality of both estimated image and depth improve when the measurements have small noise (high SNR) and the quality degrades as we add more noise in the measurements (low SNR). Another observation we can make is that the scenes that are farther away have higher RMSE. This aspect is understandable because as the scenes move farther, $\balpha$ of the scene pixels all get very close to 1 and we cannot resolve fine depth variations in the scene.

\begin{figure*}[htb!]
    \centering
    \begin{subfigure}{\figwidth}
    \caption*{Original}
    \includegraphics[width=1\linewidth]{figures/sword2_scene_image.png}
    \caption*{Image PSNR:}
    \includegraphics[width=1\linewidth]{figures/sword2_scene_depth.png}
        \caption*{Depth RMSE:}
    \end{subfigure}
    \begin{subfigure}{\figwidth}
    \caption*{25$\mu$m} 
    \includegraphics[width=1\linewidth]{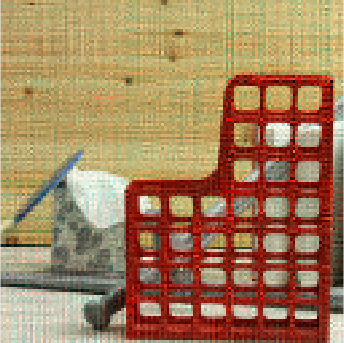}
    \caption*{24.31dB}
    \includegraphics[width=1\linewidth]{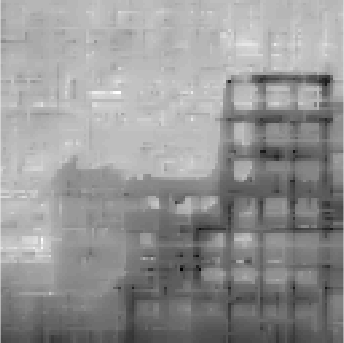}
        \caption*{45.61mm} 
    \end{subfigure}
        \begin{subfigure}{\figwidth}
    \caption*{50$\mu$m} 
    \includegraphics[width=1\linewidth]{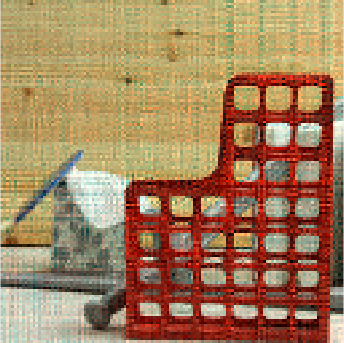}
     \caption*{22.27dB}
    \includegraphics[width=1\linewidth]{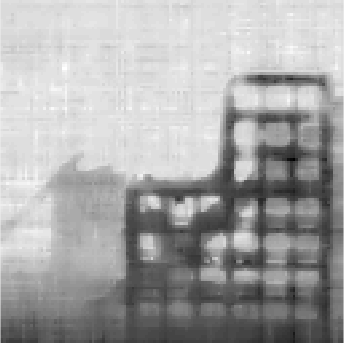}
        \caption*{34.39mm} 
    \end{subfigure}
        \begin{subfigure}{\figwidth}
    \caption*{100$\mu$m} 
    \includegraphics[width=1\linewidth]{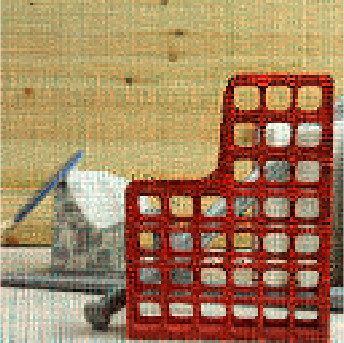}
     \caption*{21.92dB}
    \includegraphics[width=1\linewidth]{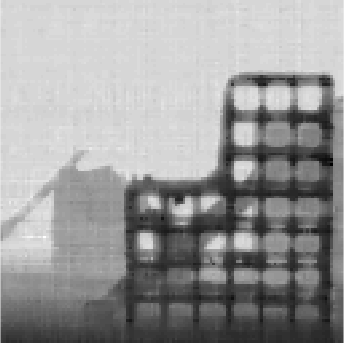}
        \caption*{24.42mm} 
    \end{subfigure}
    ~~~~
    \begin{subfigure}{\figwidth}
     \caption*{Original}
    \includegraphics[width=1\linewidth]{figures/playtable_scene_image.png}
    \caption*{Image PSNR:}
    \includegraphics[width=1\linewidth]{figures/playtable_scene_depth.png}
        \caption*{Depth RMSE:}
    \end{subfigure}
    \begin{subfigure}{\figwidth}
     \caption*{25$\mu$m}
    \includegraphics[width=1\linewidth]{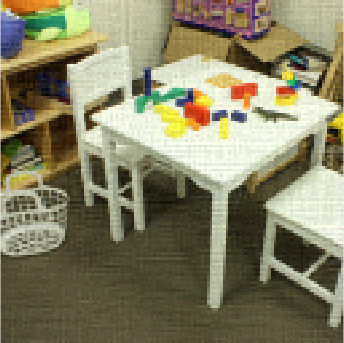}
    \caption*{32.38dB}
    \includegraphics[width=1\linewidth]{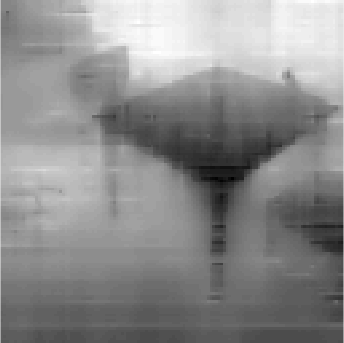}
        \caption*{198.50mm} 
    \end{subfigure}
        \begin{subfigure}{\figwidth}
         \caption*{50$\mu$m}
    \includegraphics[width=1\linewidth]{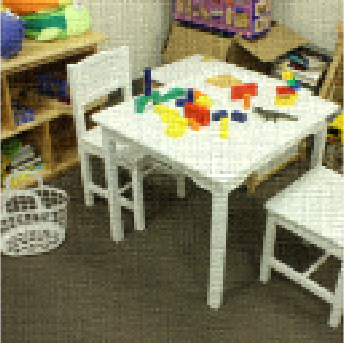}
    \caption*{31.50dB}
    \includegraphics[width=1\linewidth]{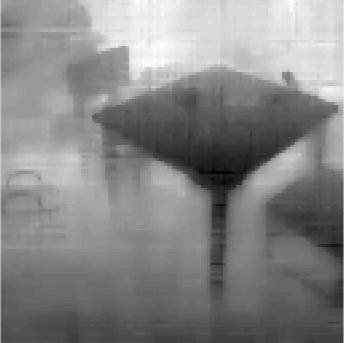}
        \caption*{147.68mm} 
    \end{subfigure}
        \begin{subfigure}{\figwidth}
         \caption*{100$\mu$m}
    \includegraphics[width=1\linewidth]{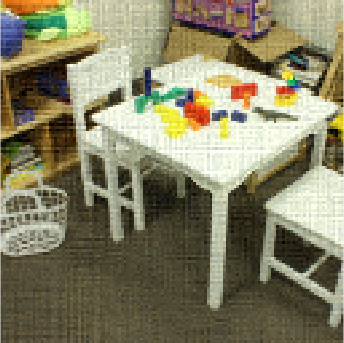}
    \caption*{28.09dB}
    \includegraphics[width=1\linewidth]{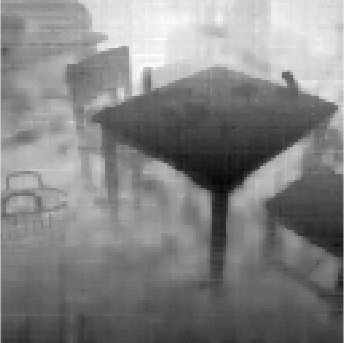}
        \caption*{134.15mm} 
    \end{subfigure}
    ~~~

    \caption{Reconstructions from measurements with different sizes of sensor pixels. The number of sensor pixels is fixed as $512\times512$. The quality of depth reconstruction improves as  we increase the size of sensor pixels.}
    \label{fig:size_of_pixels_example}
\end{figure*}

\subsection{Size of Sensor}
In conventional disparity-based depth estimation method \cite{hartley2003multiple}, the quality of reconstructed depth depends on the disparity between frames captured from multiple camera views. Larger distance between camera viewing positions results in better depth estimation accuracy. In a lensless imaging system, we can think of each pinhole on the mask and the sensor area behind the mask as a tiny pinhole camera. The analogy only goes this far, because we do not record images from these tiny pinhole cameras separately; instead, we record a multiplexed version of all the views. The disparity between different points on the sensors, however, does affect our ability to resolve the depth of the scene, which is determined by the size of sensor.

To analyze the effect of disparity in our system, we performed experiments with three different sizes of sensor pixels from 25$\mu$m, 50$\mu$m, and 100$\mu$m. For comparison, the number of sensor pixels and other parameters are set to the default settings as described earlier. No noise is included in this experiment. Results in terms of reconstructed image and depth maps are presented in Figure~\ref{fig:size_of_pixels_example}, where we observe that the quality of depth reconstruction improves as we increase the size of sensor pixels. The results in Figure~\ref{fig:size_of_pixels_example} demonstrate that increasing the disparity of viewing points increases the depth reconstruction quality.

\subsection{Comparison with Existing Methods}\label{sec:simulation_comparison}

\begin{figure*}[htb!]
\centering
    
             \begin{subfigure}{\figwidth}
             \caption*{Original}
    \includegraphics[width=1\linewidth]{figures/sword2_scene_image.png}
    \caption*{Image PSNR:}
    \includegraphics[width=1\linewidth]{figures/sword2_scene_depth.png}
        \caption*{Depth RMSE:}
    \end{subfigure}
    \begin{subfigure}{\figwidth}
    \caption*{3D Grid\cite{antipa2018diffusercam}} 
    \includegraphics[width=1\linewidth]{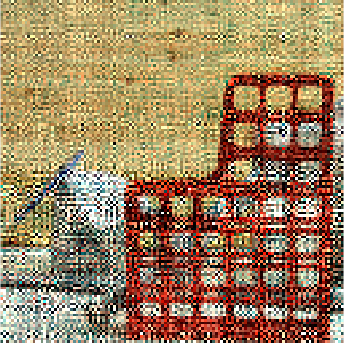}
    \caption*{9.10dB}
    \includegraphics[width=1\linewidth]{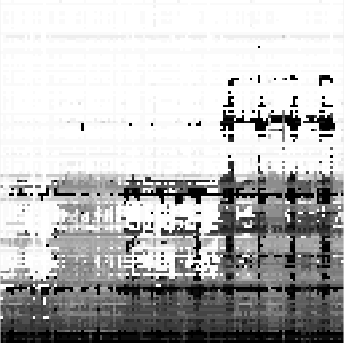}
        \caption*{87.55mm} 
    \end{subfigure}
        \begin{subfigure}{\figwidth}
        \caption*{Greedy \cite{asif2017lensless3D}}
    \includegraphics[width=1\linewidth]{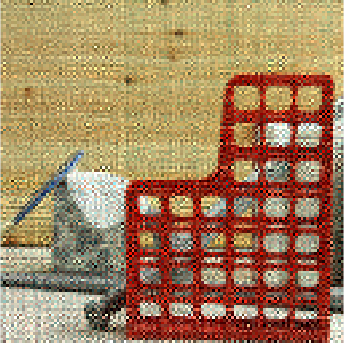}
    \caption*{14.04dB}
    \includegraphics[width=1\linewidth]{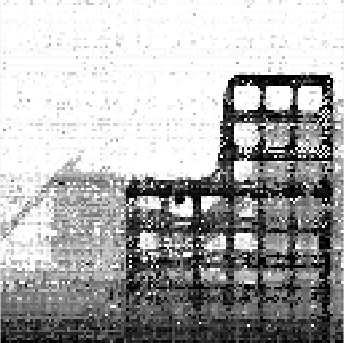}
        \caption*{47.59mm} 
    \end{subfigure}
        \begin{subfigure}{\figwidth}
        \caption*{Ours}
    \includegraphics[width=1\linewidth]{figures/sword2_pixel_size_50_image.png}
    \caption*{22.27dB}
    \includegraphics[width=1\linewidth]{figures/sword2_pixel_size_50_depth.png}
        \caption*{34.39mm} 
    \end{subfigure}
    ~~~
    \begin{subfigure}{\figwidth}
    \caption*{Original}
    \includegraphics[width=1\linewidth]{figures/playtable_scene_image.png}
    \caption*{Image PSNR:}
    \includegraphics[width=1\linewidth]{figures/playtable_scene_depth.png}
        \caption*{Depth RMSE:}
    \end{subfigure}
    \begin{subfigure}{\figwidth}
    \caption*{3D Grid\cite{antipa2018diffusercam}}
    \includegraphics[width=1\linewidth]{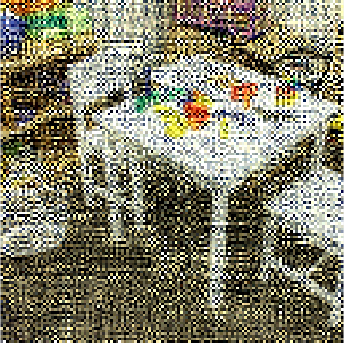}
     \caption*{7.14dB}
    \includegraphics[width=1\linewidth]{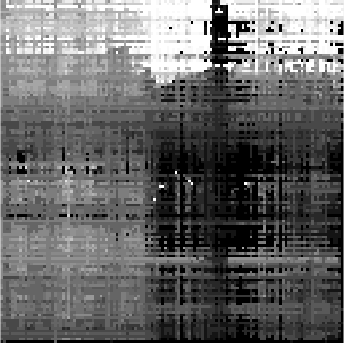}
        \caption*{479.78mm} 
    \end{subfigure}
        \begin{subfigure}{\figwidth}
        \caption*{Greedy \cite{asif2017lensless3D}}
    \includegraphics[width=1\linewidth]{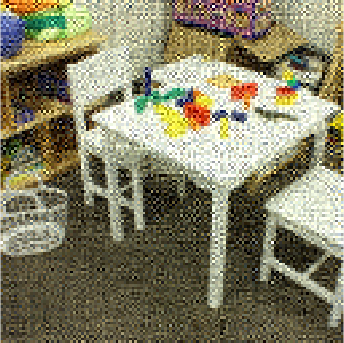}
     \caption*{16.71dB}
    \includegraphics[width=1\linewidth]{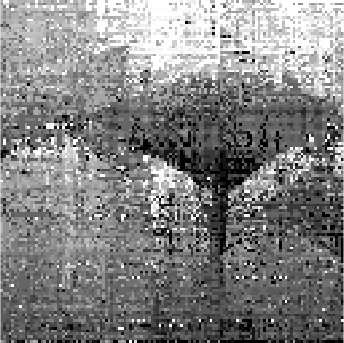}
        \caption*{400.99mm} 
    \end{subfigure}
        \begin{subfigure}{\figwidth}
        \caption*{Ours}
    \includegraphics[width=1\linewidth]{figures/playtable_pixel_size_50_image.png}
     \caption*{31.50dB}
    \includegraphics[width=1\linewidth]{figures/playtable_pixel_size_50_depth.png}
        \caption*{147.68mm} 
    \end{subfigure}
    
    \caption{Comparison of existing 3D recovery methods for lensless imaging, 3D grid method in \cite{adams2017rice_depth, antipa2018diffusercam} and greedy method in \cite{asif2017lensless3D}, with our proposed method. 3D grid method provides a 3D volume with multiple depth planes; therefore, we pick the depth with the largest light intensity along any angle for comparison.}
    \label{fig:methods_comparison}
\end{figure*}

Finally, we present a comparison of our proposed algorithm and two other methods for 3D recovery with lensless cameras. In our method, we estimate light intensity and a depth map over continuous domain. The greedy method in \cite{asif2017lensless3D} also estimates intensity and depth separately, but the depth map for any angle is restricted to one of the predetermined planes. Three-dimensional recovery using lensless cameras for 3D fluorescence microscopy was presented in \cite{antipa2018diffusercam} and \cite{adams2017rice_depth}, which estimate the entire 3D volume of the scene sampled over a predetermined 3D grid. Since the unknown volume scene in microscopy is often very sparse, the 3D scene recovery problem is solved as a sparse recovery problem for the light intensity over all the grid voxels. The result is a light distribution over the entire 3D space. We call this method 3D Grid and use the code provided in \cite{antipa2018diffusercam} to solve the 3D recovery problem using the forward model and measurements from our simulation setup. 

The imaging experiments in \cite{antipa2018diffusercam} and \cite{adams2017rice_depth} are aimed at fluorescence imaging in which objects are mostly transparent and all the points in the 3D volume can contribute to the sensor measurements without occluding one another. In contrast, we consider natural photographic scenes, where objects are usually opaque and block light from objects behind them along the same angular direction. We can model such scenes as having only one voxel along any angle to be nonzero; however, that will be a nonconvex constraint and to enforce that we will have to resort to some heuristic similar to the one in \cite{asif2017lensless3D}. 
For the sake of comparison, we solve the $\ell_1$ norm-based sparse recovery problem as described in \cite{antipa2018diffusercam}, but then we pick the points with the maximum light intensity at each angle to form the reconstructed image and (inverse) depth map. 

A comparison of different recovery methods with the same imaging setup is shown in Figure~\ref{fig:methods_comparison}. For the same scene, we reconstruct the same measurements using the three methods. As we can observe that our proposed algorithm offers a significant improvement compared to existing methods in all the test scenes.

The time and storage complexity of our proposed method and the other two methods depend on different factors; such as whether the imaging model is separable or convolutional and the sampling density along the depth. Since the main computational complexity  of all the methods arises from the applications of the forward and adjoint operators, we will just discuss the complexity of those operators for different methods. The imaging operator in the greedy algorithm uses a separable mask and assumes that the scene consists of $D$ depth planes. The computational complexity of the operator is $O(DMN^2)$ when we have $M\times M$ sensor pixels to reconstruct $N\times N$ image at $D$ predefined depth planes. The convolutional model can be implemented using a fast Fourier transform and its complexity for a 3D volume with $D$ depth planes is $O(DN^2\log(N)$. The time complexity of forward imaging operator in the proposed method is ${O}(M^2N^2)$ because we assign independent depth values to each of the angles.

\section{Experimental results}
\label{sec: experimental}
    To demonstrate the performance of our proposed method in the real world, we built a FlatCam prototype to capture images of different objects with different depth profiles. Below we discuss the details of our experiments and present reconstructed intensity and depth maps for some real objects.

\subsection{Prototype Setup} 
\noindent{\bf Image sensor.} We used a Sony IMX249 CMOS color sensor that came inside a point grey camera (model BFLY-U3-23S6C-C). The sensor has $1920\times1200$ pixels and the size of each pixel is 5.86$\mu$m. The physical size of the sensor is approximately 11.2\text{mm}$\times$7\text{mm}. 

\begin{figure}[tbh]
    \centering
    \includegraphics[width=0.49\linewidth,keepaspectratio]{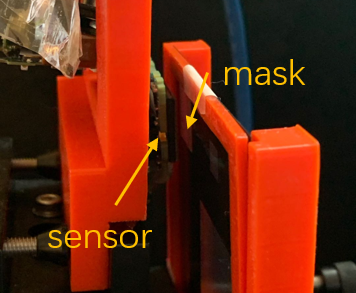}
    \includegraphics[width=0.49\linewidth,keepaspectratio]{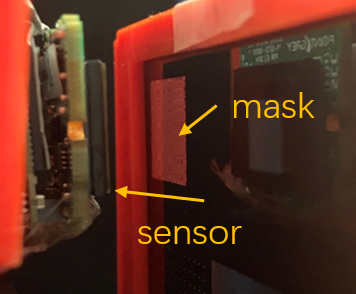}
    \caption{\textbf{Camera prototype.} The side view of the sensor and mask assembly. The sensor and mask are placed at a large distance for this image, but their distance $(d)$ is approximately 4mm in our experiments. The mask pattern is binary and separable, and the physical size of each feature is $60\mu$m.}
    \label{fig:mask_placement}
\end{figure}

\begin{figure*}[tbh]
    \centering
    \includegraphics[width=1\linewidth,keepaspectratio]{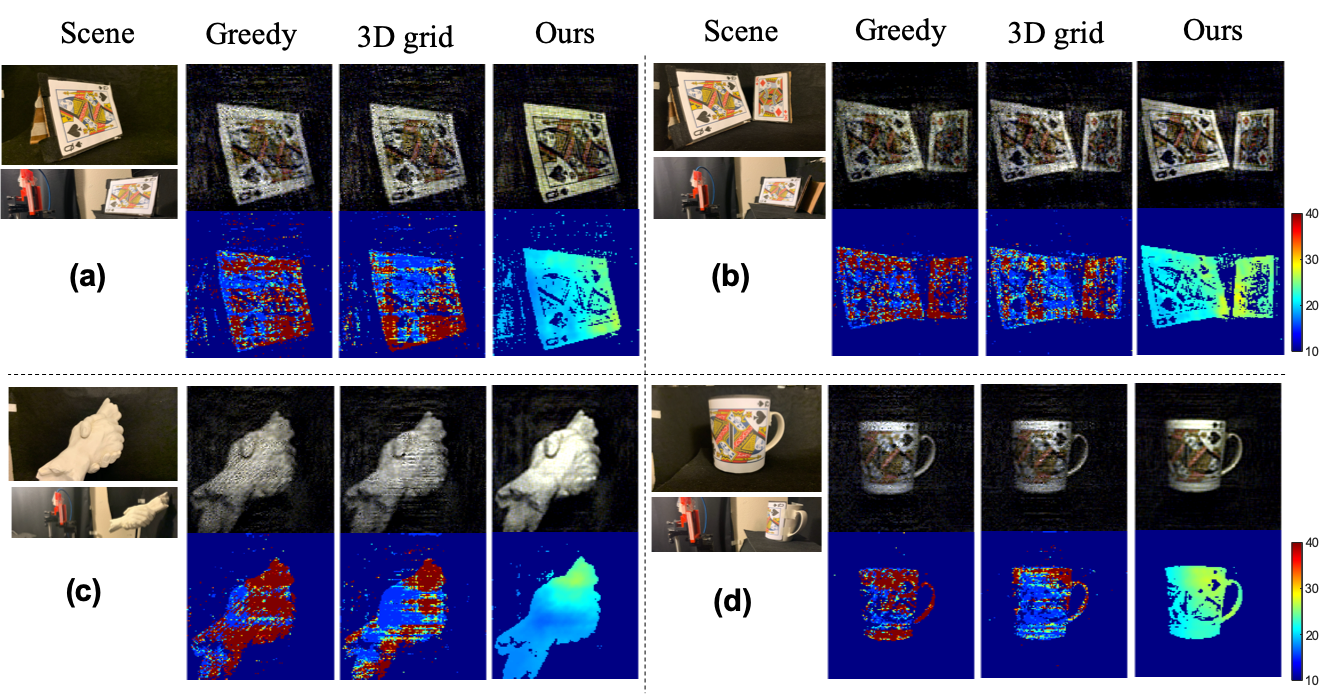}
    \caption{\textbf{Experiments on real objects.} \textbf{(a)} A slanted card; the depth range is 18--28cm \textbf{(b)} Two slanted cards; the depth range of left card is 18--28cm and the right card is 26--29cm. \textbf{(c)} Hand sculpture; depth range is 15--30cm.  \textbf{(d)} A mug with card texture; depth range is 24--27cm. We divide each group of real scenes into four columns, the first column is front view and side view of the scene, the second column is the result from greedy algorithm in  \cite{asif2017lensless3D}, the third column is the output of sparse 3D grid recovery algorithm proposed in \cite{antipa2018diffusercam} and \cite{adams2017rice_depth}, and the last column is the image intensity and depth map estimated using our proposed algorithm. }
    \label{fig:real_scenes}
\end{figure*}

\noindent{\bf Mask pattern.} We printed a binary mask pattern on a plastic sheet. The mask pattern was created by computing an outer product of two 255-length MLS vectors and setting all the -1 entries to 0. The physical size of each mask feature is 60$\mu$m. The physical size of the generated mask pattern is approximately 15.3mm$\times$15.3mm. 

\noindent{\bf Sensor and mask placement.} We placed the mask and the bare sensor on two optical posts such that the mask-to-sensor distance $(d)$ is approximately 4mm; we attached kinematic platforms on top of the optical posts so that we can align the sensor and mask. Pictures of our sensor and mask setup are shown in Figure~\ref{fig:mask_placement}.

\noindent{\bf Data acquisition and processing.} In our experiments, we calibrated the system by capturing sensor measurements while moving an LED flashlight at different locations in front of the camera. We performed all our experiments by uniformly illuminating the object with a table lamp. 
We reconstructed depth map and colored images at $128\times128$ pixel resolution from $512\times512$ sensor measurements. The sensor provides $1920\times 1200$ pixels; we first resize the sensor measurement into $960\times 600$ pixels by binning $2\times 2$ pairs, and then we crop a $512 \times512$ area in the center.

\subsection{Calibration of the Prototype Camera}
\label{calibrate_prototype}
We use a separable mask pattern and align the mask and sensor assembly such that the response of any point source on the sensor is a rank-one image after mean subtraction \cite{asif2017flatcam}. 
To calibrate system matrix at one given depth, we can capture a sequence of rank-1 Hadamard patterns as described in \cite{asif2017flatcam} or capture the response of one LED flashlight as described in \cite{antipa2018diffusercam}. 
Instead of calibrating the separable system matrices for different depth planes, we calibrated the mask pattern function at one depth and evaluated the point spread function at arbitrary depth and angle according to  \eqref{eq:maskFunction}. 
Because our mask pattern is bigger than the sensor, we captured sensor measurements for LED flashlight at 9 different angles at the same depth and merged them to estimate the mask function at that depth. 

In our experiments, we captured the sensor measurements by placing an LED at $z=42cm$ away from the sensor, which corresponds to the mask function in \eqref{eq:maskFunction} evaluated at $\alpha = 1-d/z =0.9905$ for $d=4mm, z=42cm$. We first resized the calibrated mask function to compute the mask function corresponding to $\alpha =1$. 

\subsection{Reconstruction of Real Objects}
We present results for four objects in Figure~\ref{fig:real_scenes}, (a) slanted card has depth range from 18cm to 28cm, (b) two slanted cards have depth ranges from 18cm to 28cm and 26cm to 29cm, (c) hands sculpture has depth range from 15cm to 30cm, and (d) mug with card texture depth is from 24cm to 27cm. The figure is divided into four boxes. In each box, we present a front- and side-view of the object along with estimated scene intensity and depth maps for three methods. the greedy algorithm in  \cite{asif2017lensless3D}, the sparse 3D volume recovery method from \cite{adams2017rice_depth, antipa2018diffusercam}, and our proposed method. For the greedy and 3D grid method, we generated 15 candidate depth planes by uniformly sampling the inverse depth values $\alpha$ between 0.96 and 0.9905 (corresponding to the depth of 10cm and 42cm, respectively).

All the objects in our experiments are placed in front of the black background and the depth values for dark pixels are not meaningful. We can observe that in all these experiments, our proposed method provides a continuous depth map that is consistent with the real depth of the object in the scene. In comparison, both the greedy algorithm \cite{asif2017lensless3D} and the sparse 3D volume recovery algorithm \cite{adams2017rice_depth, antipa2018diffusercam} produce coarse and discretized depth maps.
The intensity map recovered by our method is also visually better compared to other methods. 

Even though our proposed algorithm produces better intensity and depth maps compared to the greedy and 3D grid methods, we observed that the estimated depth has some errors in the darker parts of the objects. For instance, the left part of the mug is darker than the right part because the object was illuminated from a lamp on the right side. The left part appears to have errors in the depth estimate as several pixels are assigned small depth values, but that part is in fact farther from the sensor. We also observe a similar effect in other experiments, where depth estimates for darker parts of the scene appear to have larger errors.


\section{Conclusion}
We presented a new algorithm to jointly estimate the image and depth of a scene using a single snapshot of a mask-based lensless camera. Existing methods for 3D lensless imaging either estimate scene over a predefined 3D grid (which is computationally expensive) or a small number of candidate depth planes (which provides a coarse depth map). We divide the scene into an intensity map at uniform angles and a depth map on a continuous domain, which allows us to estimate a variety of scenes with different depth ranges using the same formulation. We jointly estimate the image intensity and depth map by solving a nonconvex problem. We initialize our estimates using a greedy method and add weighted regularization to enforce smoothness in the depth estimate while preserving the sharp edges. We demonstrated with extensive simulations and experiments with real data that our proposed method can recover image and depth with high accuracy for a variety of scenes. We evaluated the performance of our methods under different noise levels, sensor sizes, and numbers of sensor pixels and found the method to be robust. We presented a comparison with existing methods for lensless 3D imaging and demonstrated both in simulation and real experiments that our method provides significantly better results. We believe this work provides a step toward capturing complex scenes with lensless cameras, where depth estimation is a feature as well as a compulsion because if the depth information is unavailable or inaccurate, that will cause artifacts in the recovered images.

\bibliographystyle{IEEEtran}
\bibliography{refs}


\end{document}